\documentclass[submission, copyright, creativecommons]{eptcs}
\providecommand{\event}{Applied Category Theory 2021 (ACT 2021)}

\usepackage{amsmath}
\usepackage{amssymb}
\usepackage{graphicx}
\usepackage{graphbox}
\usepackage{tikz-cd}
\usepackage{proof}

\newtheorem{definition}{Definition}

\title{Fast Automated Reasoning over String Diagrams using Multiway Causal Structure}
\author{Jonathan Gorard\institute{University of Cambridge,\\Cambridge, United Kingdom}\email{jg865@cam.ac.uk}
\and Manojna Namuduri\institute{Wolfram Research,\\Illinois, USA}\email{manon@wolfram.com}
\and Xerxes D. Arsiwalla\institute{Pompeu Fabra University,\\Barcelona, Spain}\email{x.d.arsiwalla@gmail.com}}
\def\titlerunning{Fast Automated Reasoning over String Diagrams using Multiway Causal Structure}
\def\authorrunning{J. Gorard, M. Namuduri \& X. D. Arsiwalla}

\begin{document}
\maketitle

\begin{abstract}
We introduce an intuitive algorithmic methodology for enacting automated rewriting of string diagrams within a general double-pushout (DPO) framework, in which the sequence of rewrites is chosen in accordance with the causal structure of the underlying diagrammatic calculus. The combination of the rewriting structure and the causal structure may be elegantly formulated as a weak 2-category equipped with both total and partial monoidal bifunctors, thus providing a categorical semantics for the full \textit{multiway evolution causal graph} of a generic \textit{Wolfram model} hypergraph rewriting system. As an illustrative example, we show how a special case of this algorithm enables highly efficient automated simplification of quantum circuits, as represented in the ZX-calculus.
\end{abstract}

\section{Introduction}

String diagrams, as first formalized by Joyal and Street\cite{Joyal1991}, provide a rigorous and elegant graphical language for representing arbitrary monoidal categories. Efficient reasoning over string diagrams is consequently a mainstay of many areas of applied category theory, including the ZX-calculus\cite{Coecke2008}\cite{Coecke2009a} and categorical quantum mechanics\cite{Abramsky2004}\cite{Abramsky2009}, concurrency theory\cite{Bonchi2014}, network and control theory\cite{Baez2014}\cite{Baez2017}\cite{Baez2018}, and even computational linguistics\cite{Coecke2010a}\cite{Bolt2017}; it is therefore unsurprising that a variety of popular interactive proof assistants for string diagrams, such as PyZX\cite{Kissinger2019} for the ZX-calculus and Quantomatic\cite{Kissinger2015} for more general diagrammatic calculi, have previously been developed. Such automated reasoning systems over string diagrams constitute an important special case of \textit{Wolfram model multiway systems}\cite{Wolfram2002a}\cite{Wolfram2020}\cite{Gorard2020}\cite{Gorard2020d}, which describe abstract rewriting systems over arbitrary (hyper)graphs, as well as their causal structure, in terms of double-pushout rewriting systems over partial adhesive categories\cite{Gorard2020b}\cite{Gorard2021a}. The purpose of the present paper is to introduce a new algorithmic framework for enacting fast automated reasoning over string diagrams, in which the lemmas introduced by the automated theorem-proving system are selected on the basis of their causal structure, thereby maximizing the total number of outgoing causal edges associated with the corresponding path in the multiway system, and argue that this is an effective heuristic for selecting those lemmas which are most likely to exert the greatest effect on the shortening of subsequent proofs.

In Section \ref{sec:Section1}, we show how the Wolfram model can be formulated as a double-pushout (DPO) rewriting system over string diagrams for hypergraph categories, and discuss how the concurrency and parallelism theorems of algebraic graph transformation theory can be used to extract a symmetric monoidal category structure from such a rewriting system. We also discuss how the formalism of \textit{causal categories} due to Coecke and Lal\cite{Coecke2013} also permits a compositional description of the causal structure of such a rewriting system, allowing the entire Wolfram model multiway evolution causal graph (combining both the rewriting structure and the causal structure) to be formulated in terms of a weak 2-category equipped with a total monoidal bifunctor on the 1-cells and a partial monoidal bifunctor on the 2-cells. In Section \ref{sec:Section2}, we proceed to show how this weak 2-category structure allows us to extend the traditional (unfailing) Knuth-Bendix completion algorithm to produce a refutation-complete proof calculus that accounts for the causal structure of the underlying diagrammatic reasoning language. Finally, in Section \ref{sec:Section3}, we illustrate an application of this general algorithm in a notable special case, namely the problem of diagrammatic simplification of quantum circuits in the ZX-calculus, by presenting a fully worked example of an automatically-generated proof of unitarity for the CNOT gate.

All of the code necessary to reproduce the computations presented within this paper is open source and available for free (with extensive documentation) on the \textit{Wolfram Function Repository}. For instance \href{https://resources.wolframcloud.com/FunctionRepository/resources/MakeZXDiagram}{MakeZXDiagram} enables one to construct string diagrams for quantum circuits; \href{https://resources.wolframcloud.com/FunctionRepository/resources/QuantumDiscreteStateToZXDiagram/}{QuantumDiscreteStateToZXDiagram} / \href{https://resources.wolframcloud.com/FunctionRepository/resources/ZXDiagramToQuantumDiscreteState/}{ZXDiagramToQuantumDiscreteState} enable one to convert ZX-diagrams to and from purely symbolic quantum objects in the Wolfram Language's open source quantum computing framework; \href{https://resources.wolframcloud.com/FunctionRepository/resources/MultiwayOperatorSystem}{MultiwayOperatorSystem} allows one to evolve the resulting multiway systems and extract their causal structure; \href{https://resources.wolframcloud.com/FunctionRepository/resources/FindWolframModelProof}{FindWolframModelProof} allows one to construct automatic proofs of equivalence between (hypergraph) string diagrams; etc.

\section{The Compositional Structure of (Hyper)graph Rewriting Systems}
\label{sec:Section1}

Whereas string diagrams for ordinary monoidal categories correspond to (directed) graphs, string diagrams for \textit{hypergraph categories} (in the terminology of Kissinger\cite{Kissinger2014} and Fong\cite{Fong2015}\cite{Fong2016}) correspond to hypergraphs.

\begin{definition}
A ``hypergraph category'' is a symmetric monoidal category ${\left( \mathbf{C}, \otimes , I \right)}$ in which every object $A$ in ${\mathrm{ob} \left( \mathbf{C} \right)}$ is equipped with a special commutative Frobenius algebra structure ${\left( A, \mu, \eta, \delta, \epsilon \right)}$, such that the Frobenius algebra structure of the monoidal product ${A \otimes B}$ (for $A$ and $B$ in ${\mathrm{ob} \left( \mathbf{C} \right)}$) is canonically induced from the Frobenius algebra structures of $A$ and $B$.
\end{definition}
A \textit{typed hypergraph production}\cite{Ehrig1973}\cite{Ehrig2006} is now a span of monomorphisms $p$:

\begin{equation}
p = \left( \begin{tikzcd}
L & K \arrow[l, "l"] \arrow[r, "r"] & R
\end{tikzcd} \right),
\end{equation}
with $L$, $K$ and $R$ being typed hypergraphs, and $l$ and $r$ being injective typed hypergraph morphisms. A \textit{direct typed hypergraph transformation} ${G \Rightarrow^{p, m} H}$ is then given by the following pair of pushout squares:

\begin{equation}
\begin{tikzcd}
L \arrow[d, "m"] & K \arrow[l, "l"] \arrow[d, "k"] \arrow[r, "r"] & R \arrow[d, "n"]\\
G & D \arrow[l, "f"] \arrow[r, "g"] & H
\end{tikzcd},
\end{equation}
assuming production $p$ and match ${m : L \to G}$. A \textit{typed hypergraph transformation} ${G_0 \Rightarrow^{*} G_n}$ is thus any sequence ${G_0 \Rightarrow G_1 \Rightarrow \dots \Rightarrow G_n}$ of direct typed hypergraph transformations.

Such double-pushout (DPO) rewriting systems are conventionally defined over \textit{adhesive categories}\cite{Lack2004a}, i.e. over categories that have \textit{pushouts along monomorphisms}, that have pullbacks, and in which every pushout along a monomorphism satisfies the \textit{van-Kampen square} condition. A \textit{pushout along a monomorphism} is any pushout square:

\begin{equation}
\begin{tikzcd}
A \arrow[r, "g"] \arrow[d, "f"] & B \arrow[d, "f^{\prime}"]\\
C \arrow[r, "g^{\prime}"] & D
\end{tikzcd},
\end{equation}
of a span of the form:

\begin{equation}
\begin{tikzcd}
B & A \arrow[l, "g"] \arrow[r, "f"] & C
\end{tikzcd},
\end{equation}
in which either $f$ or $g$ is a monomorphism. Such a pushout square is said to satisfy the \textit{van-Kampen square condition} if and only if, for every commutative diagram of the form:

\begin{equation}
\begin{tikzcd}
B^{\prime} \arrow[ddd, "f_{h}^{\prime}"] \arrow[dr, "h_B"] & & & A^{\prime} \arrow[lll, "g_h"] \arrow[dl, "h_A"] \arrow[ddd, "f_h"]\\
& B \arrow[d, "f^{\prime}"] & A \arrow[l, "g"] \arrow[d, "f"] &\\
& D & C \arrow[l, "g^{\prime}"] &\\
D^{\prime} \arrow[ur, "h_D"] & & & C^{\prime} \arrow[lll, "g_{h}^{\prime}"] \arrow[ul, "h_C"]
\end{tikzcd},
\end{equation}
within which the subdiagrams:

\begin{equation}
\begin{tikzcd}
B^{\prime} \arrow[d, "h_B"] & A^{\prime} \arrow[l, "g_h"] \arrow[d, "h_A"]\\
B & A \arrow[l ,"g"]
\end{tikzcd}, \qquad \text{ and } \qquad
\begin{tikzcd}
A \arrow[d, "f"] & A^{\prime} \arrow[l, "h_A"] \arrow[d, "f_h"]\\
C & C^{\prime} \arrow[l, "h_C"]
\end{tikzcd},
\end{equation}
are pullbacks, the pushouts and pullbacks are compatible. However, the category of typed hypergraphs that we consider here is not fully adhesive, since the arbitrary nature of the connectivity of hypergraph vertices implies that pushouts along monomorphisms are not always guaranteed to exist. Nevertheless, the category of Wolfram model hypergraphs is still \textit{partial adhesive}\cite{Kissinger2011} in the sense that it forms a full subcategory ${\mathbf{C}^{\prime}}$ of an adhesive category ${\mathbf{C}}$ for which the embedding functor ${S : \mathbf{C}^{\prime} \to \mathbf{C}}$ preserves monomorphisms, which is sufficient for the definition of a DPO rewriting system\cite{Gorard2021a}.

Such a rewriting system naturally gives rise to a symmetric monoidal category ${\textbf{MuGraph}}$\cite{Gorard2020b}, representing the compositional structure of a \textit{multiway evolution graph}. The ordinary composition of morphisms in ${\textbf{MuGraph}}$ arises from the fact that productions ${p_1}$ and ${p_2}$, yielding an $E$-related transformation sequence ${G \Rightarrow H \Rightarrow G^{\prime}}$, can be composed by means of the \textit{concurrency theorem} in algebraic graph transformation theory\cite{Ehrig2006} to obtain the $E$-concurrent production ${p_1 *_{E} p_2}$, yielding instead the direct transformation ${G \Rightarrow G^{\prime}}$:

\begin{equation}
\begin{tikzcd}
& H \arrow[Rightarrow, "p_2"]{dr} & \\
G \arrow[Rightarrow, "p_1"]{ur} \arrow[Rightarrow, "p_1 *_{E} p_2"]{rr} & & G^{\prime}
\end{tikzcd}.
\end{equation}
Likewise, the monoidal composition of morphisms in ${\textbf{MuGraph}}$ arises from the fact that productions ${p_1}$ and ${p_2}$, yielding sequentially-independent transformation sequences ${G \Rightarrow H_1 \Rightarrow G^{\prime}}$ and ${G \Rightarrow H_2 \Rightarrow G^{\prime}}$, can be composed by means of the \textit{parallelism theorem}\cite{Ehrig2006} to obtain the parallel production ${p_1 + p_2}$, yielding instead the direct transformation ${G \Rightarrow G^{\prime}}$:

\begin{equation}
\begin{tikzcd}
& G \arrow[Rightarrow, "p_1"]{dl} \arrow[Rightarrow, "p_1 + p_2"]{dd} \arrow[Rightarrow, "p_2"]{dr} &\\
H_1 \arrow[Rightarrow, "p_2"]{dr} & & H_2 \arrow[Rightarrow, "p_1"]{dl}\\
& G^{\prime} &
\end{tikzcd}.
\end{equation}

\begin{figure}[ht]
\centering
\includegraphics[align=c, width=0.205\textwidth]{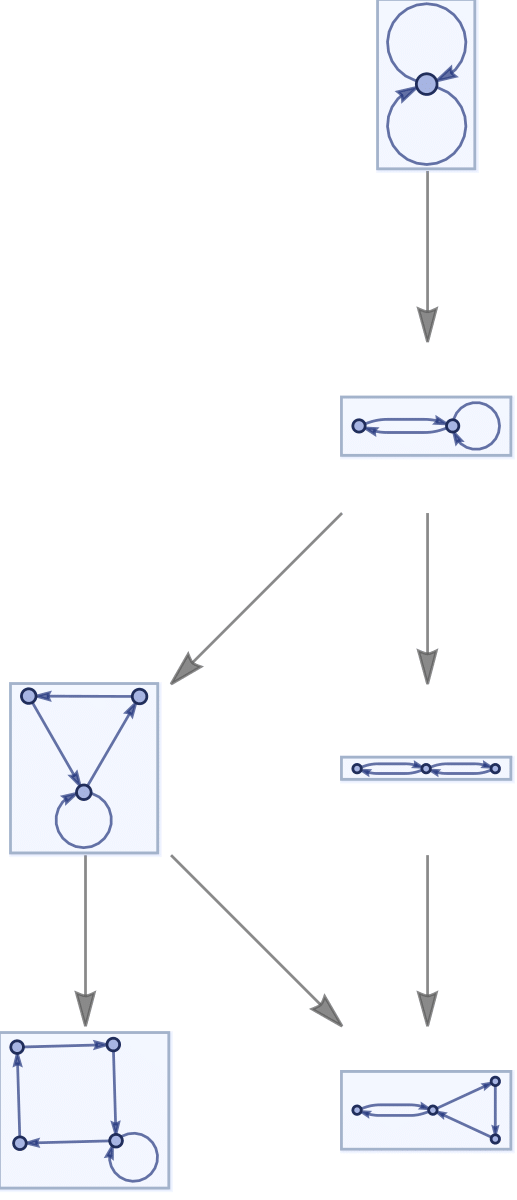}\hspace{0.3\textwidth}
\includegraphics[align=c, width=0.295\textwidth]{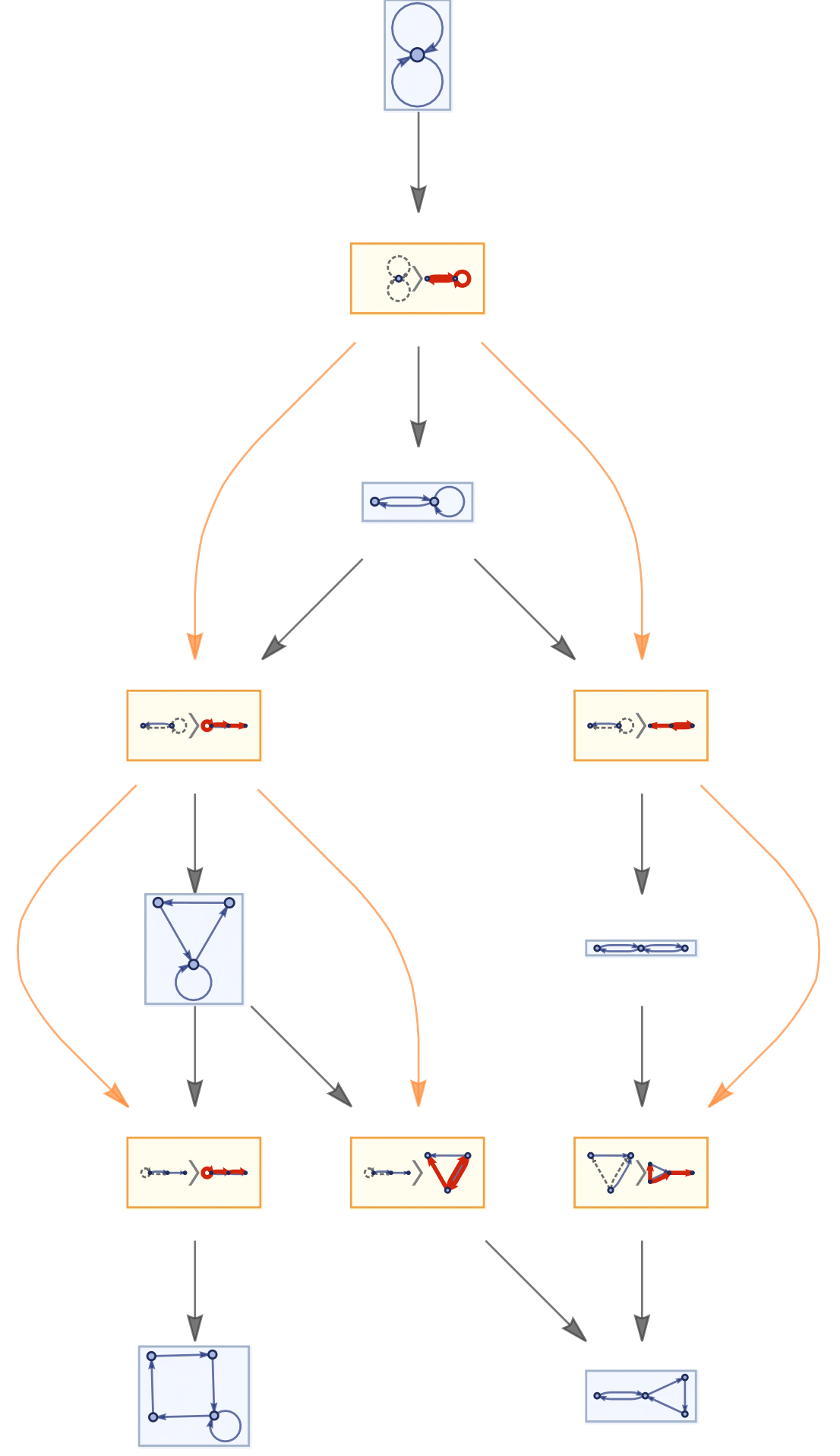}
\caption{On the left, the multiway evolution graph corresponding to the first 3 steps in the non-deterministic evolution history for the hypergraph substitution rule ${\left\lbrace \left\lbrace x, y \right\rbrace, \left\lbrace x, z \right\rbrace \right\rbrace \to \left\lbrace \left\lbrace x, z \right\rbrace, \left\lbrace x, w \right\rbrace, \left\lbrace w, y \right\rbrace \right\rbrace}$, illustrating the symmetric monoidal category structure of ${\textbf{MuGraph}}$. On the right, the corresponding multiway evolution causal graph (with productions shown as yellow vertices, and with 2-cells shown as orange edges), illustrating the weak 2-category structure of ${\textbf{MuCauGraph}}$.}
\label{fig:Figure1}
\end{figure}

Moreover, the productions themselves can exhibit causal dependencies upon one another. More specifically, productions ${p_1}$ and ${p_2}$:

\begin{equation}
p_1 = \left( \begin{tikzcd}
L_1 & K_1 \arrow[l, "l_1"] \arrow[r, "r_1"] & R_1
\end{tikzcd} \right), \qquad p_2 = \left( \begin{tikzcd}
L_2 & K_2 \arrow[l, "l_2"] \arrow[r, "r_2"] & R_2
\end{tikzcd} \right),
\end{equation}
may be said to be \textit{causally related} if the ``input'' of production ${p_2}$ (i.e. ${L_2 \setminus K_2}$) makes use of (hyper)edges that appear in the ``output'' of production ${p_1}$ (i.e. ${R_1 \setminus K_1}$)\cite{Gorard2020c}\cite{Gorard2021b}:

\begin{equation}
\left( L_2 \setminus K_2 \right) \cap \left( R_1 \setminus K_1 \right) \neq \emptyset.
\end{equation}
These causal relationships between productions thus form 2-cells within the category ${\textbf{MuGraph}}$, yielding a weak 2-category\cite{Benabou1967} that we shall henceforth call ${\textbf{MuCauGraph}}$, representing the compositional structure of a \textit{multiway evolution causal graph}\cite{Gorard2021a}. In Figure \ref{fig:Figure1}, we illustrate the symmetric monoidal category structure of ${\textbf{MuGraph}}$ by means of an explicit multiway evolution graph, in which all productions are shown as directed edges, in addition to the weak 2-category structure of ${\textbf{MuCauGraph}}$ by means of an explicit multiway evolution causal graph, in which all productions are shown as yellow vertices, with 2-cells (causal relationships) shown as orange edges. In all such multiway systems, state vertices are merged on the basis of hypergraph isomorphism, using a generalization of the algorithm presented in \cite{Gorard2016}.

Clearly, these causal edges may also be composed either in sequence or in parallel, implying that ${\textbf{MuCauGraph}}$ is in fact equipped with two distinct monoidal structures: ${\otimes_{M}}$, arising from the parallel composition of productions (1-cells), and ${\otimes_{C}}$, arising from the parallel composition of causal relationships between productions (2-cells). However, the causal monoidal structure ${\otimes_{C}}$ is only a \textit{partial} monoidal structure in the sense of Coecke and Lal\cite{Coecke2013}, i.e. if ${\textbf{CauGraph}}$ denotes the category such that ${\mathrm{ob} \left( \textbf{CauGraph} \right)}$ is the set of productions and ${\mathrm{hom} \left( \textbf{CauGraph} \right)}$ is the set of causal relationships between productions, then:

\begin{equation}
\otimes_{C} : \textbf{CauGraph} \times \textbf{CauGraph} \to \textbf{CauGraph},
\end{equation}
is a \textit{partial bifunctor}, namely a bifunctor:

\begin{equation}
\overline{\otimes_C} : \overline{\textbf{CauGraph}} \times \overline{\textbf{CauGraph}} \to \textbf{CauGraph},
\end{equation}
where ${\overline{\textbf{CauGraph}} \times \overline{\textbf{CauGraph}}}$ denotes a subcategory of ${\textbf{CauGraph} \times \textbf{CauGraph}}$ (known as the \textit{domain of definition} of ${\otimes_C}$, denoted ${\mathrm{dd} \left( \otimes_C \right)}$). Applying again the terminology of Coecke and Lal\cite{Coecke2013}, if the partial monoidal category ${\left( \textbf{CauGraph}, \otimes_C, I \right)}$ is symmetric strict, if every object $A$ in ${\mathrm{ob} \left( \textbf{CauGraph} \right)}$ contains at least one element:

\begin{equation}
\textbf{CauGraph} \left( I, A \right) \neq \emptyset,
\end{equation}
if the unit object $I$ in ${\mathrm{ob} \left( \textbf{CauGraph} \right)}$ is terminal, with unique morphism ${\top_A}$ for each object $A$ in ${\mathrm{ob} \left( \textbf{CauGraph} \right)}$:

\begin{equation}
\top_A : A \to I,
\end{equation}
and finally if the monoidal product ${A \otimes_C B}$ exists (for objects $A$ and $B$ in ${\mathrm{ob} \left( \textbf{CauGraph} \right)}$) if and only if:

\begin{equation}
\textbf{CauGraph} \left( A, B \right) = \left[ \textbf{CauGraph} \left( I, B \right) \right] \circ \top_A, \qquad \textbf{CauGraph} \left( B, A \right) = \left[ \textbf{CauGraph} \left( I, A \right) \right] \circ \top_B,
\end{equation}
then ${\textbf{CauGraph}}$ indeed forms a \textit{causal category}, as required. The partial nature of the monoidal structure ${\otimes_C}$ arises from the fact that causal edges may only be composed in parallel if the corresponding productions are causally independent.

\section{Diagrammatic Theorem-Proving using Multiway Causal Structure}
\label{sec:Section2}

\begin{figure}[ht]
\centering
\includegraphics[width=0.295\textwidth]{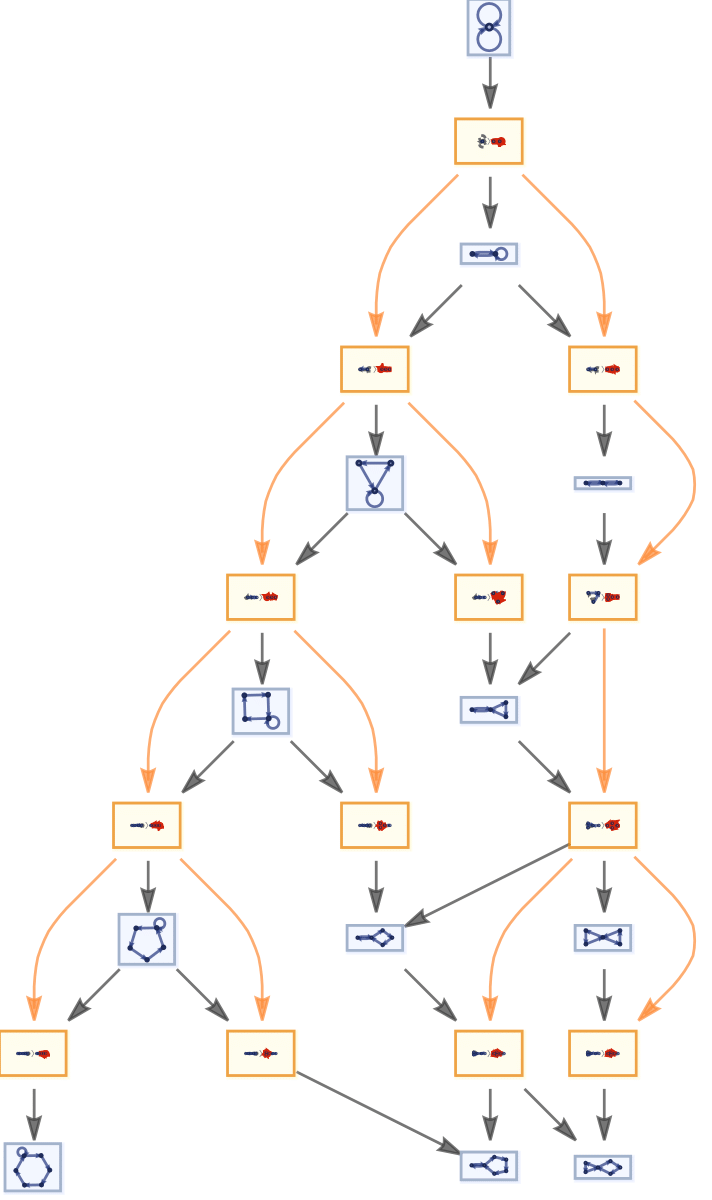}\hspace{0.2\textwidth}
\includegraphics[width=0.295\textwidth]{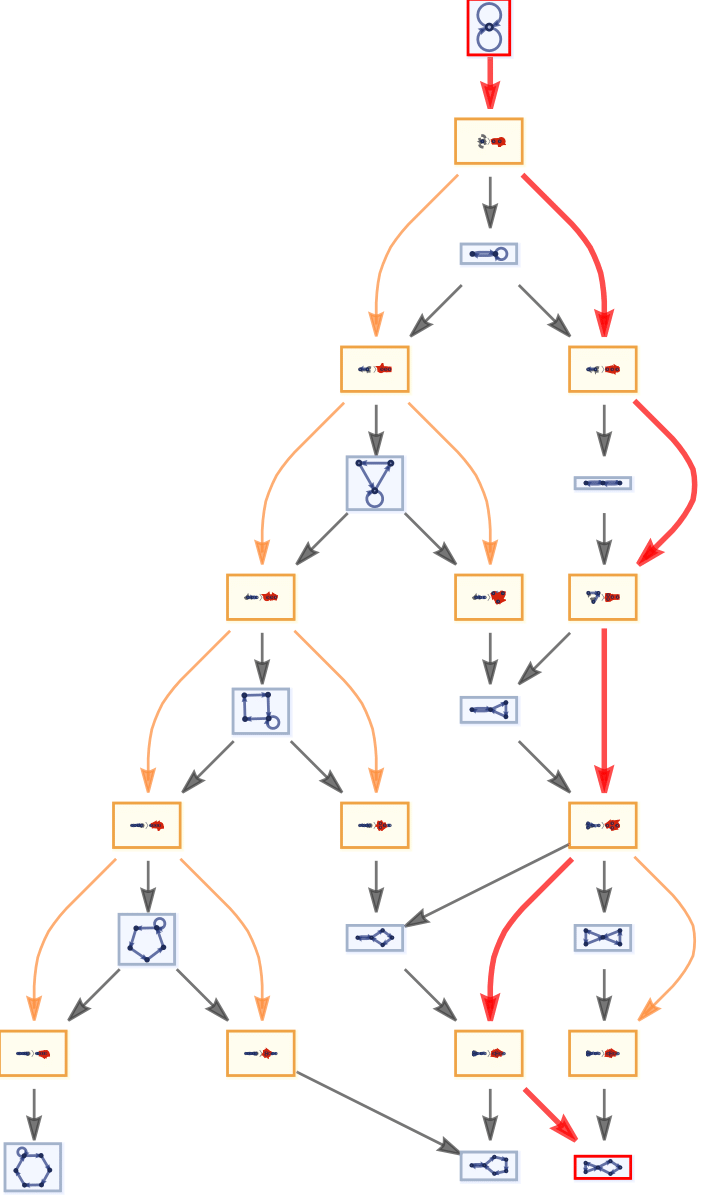}
\caption{On the left, the multiway evolution causal graph corresponding to the first 5 steps in the non-deterministic evolution history for the hypergraph substitution rule ${\left\lbrace \left\lbrace x, y \right\rbrace, \left\lbrace x, z \right\rbrace \right\rbrace \to \left\lbrace \left\lbrace x, z \right\rbrace, \left\lbrace x, w \right\rbrace, \left\lbrace w, y \right\rbrace \right\rbrace}$. On the right, the highlighted path between hypergraph state vertices ${\left\lbrace \left\lbrace 0, 0 \right\rbrace, \left\lbrace 0, 0 \right\rbrace \right\rbrace}$ and ${\left\lbrace \left\lbrace 0, 1 \right\rbrace, \left\lbrace 1, 2 \right\rbrace, \left\lbrace 2, 0 \right\rbrace, \left\lbrace 0, 3 \right\rbrace, \left\lbrace 3, 4 \right\rbrace, \left\lbrace 4, 5 \right\rbrace, \left\lbrace 5, 0 \right\rbrace \right\rbrace}$.}
\label{fig:Figure2}
\end{figure}

\begin{figure}[ht]
\centering
\includegraphics[width=0.695\textwidth]{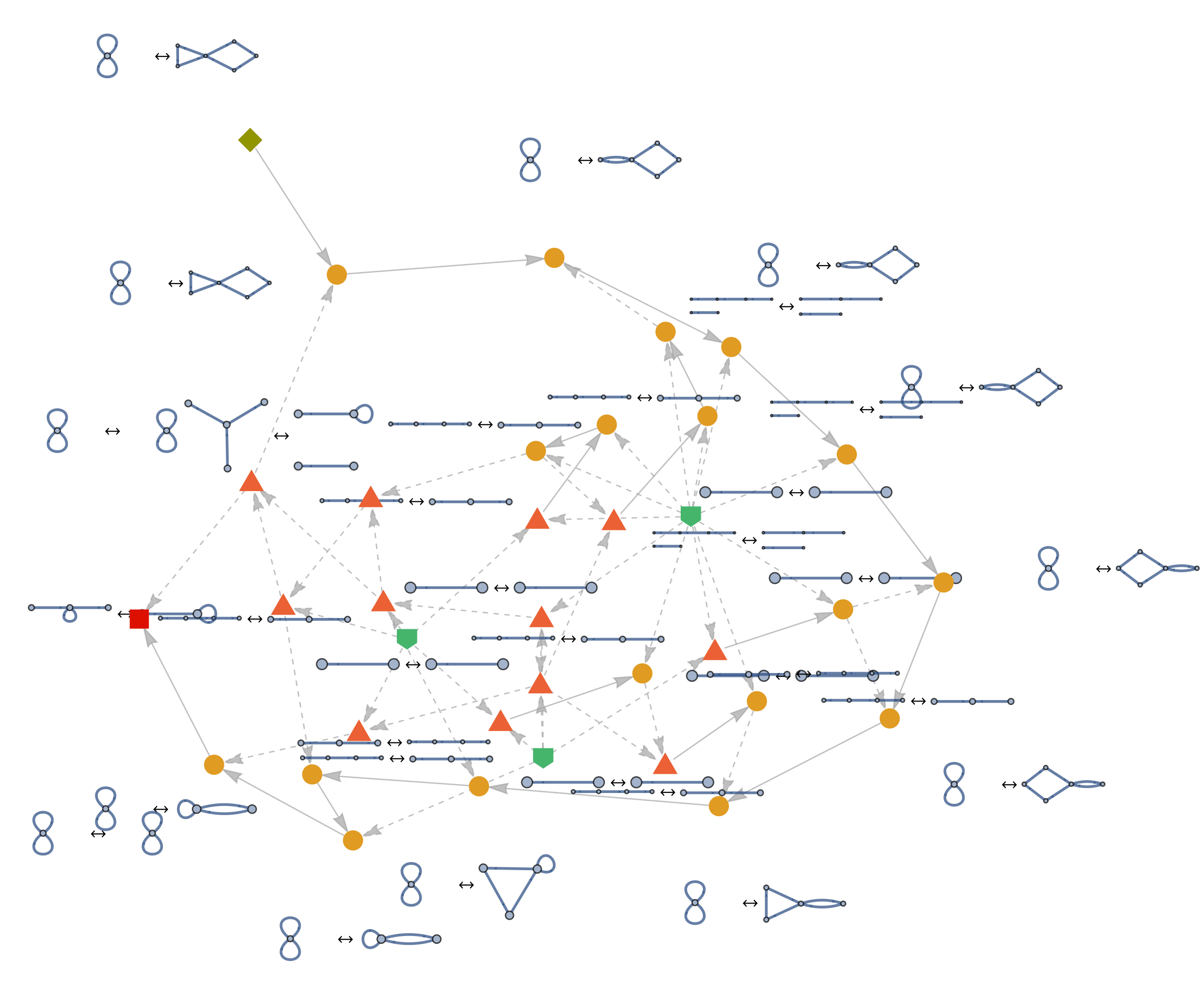}
\caption{The proof graph corresponding to the proof of the proposition that the hypergraph state ${\left\lbrace \left\lbrace 0, 1 \right\rbrace, \left\lbrace 1, 2 \right\rbrace, \left\lbrace 2, 0 \right\rbrace, \left\lbrace 0, 3 \right\rbrace, \left\lbrace 3, 4 \right\rbrace, \left\lbrace 4, 5 \right\rbrace, \left\lbrace 5, 0 \right\rbrace \right\rbrace}$ is reachable from the initial hypergraph state ${\left\lbrace \left\lbrace 0, 0 \right\rbrace, \left\lbrace 0, 0 \right\rbrace \right\rbrace}$, subject to the hypergraph substitution rule ${\left\lbrace \left\lbrace x, y \right\rbrace, \left\lbrace x, z \right\rbrace \right\rbrace \to \left\lbrace \left\lbrace x, z \right\rbrace, \left\lbrace x, w \right\rbrace, \left\lbrace w, y \right\rbrace \right\rbrace}$. Here, pointed light green boxes represent axioms, dark orange triangles represent critical pair lemmas (i.e. instances of completions/superpositions/paramodulations), light orange circles represent substitution lemmas (i.e. instances of resolutions/factorings), and dark green diamonds represent hypotheses. Solid lines represent substitutions, and dashed lines represent derived inference rules.}
\label{fig:Figure3}
\end{figure}

In Figure \ref{fig:Figure2} we show an example of a path through a multiway evolution causal graph corresponding to the proof that a hypergraph isomorphic to ${\left\lbrace \left\lbrace 0, 1 \right\rbrace, \left\lbrace 1, 2 \right\rbrace, \left\lbrace 2, 0 \right\rbrace, \left\lbrace 0, 3 \right\rbrace, \left\lbrace 3, 4 \right\rbrace, \left\lbrace 4, 5 \right\rbrace, \left\lbrace 5, 0 \right\rbrace \right\rbrace}$ is reachable from a double self-loop initial condition ${\left\lbrace \left\lbrace 0, 0 \right\rbrace, \left\lbrace 0, 0 \right\rbrace \right\rbrace}$, when subject to the hypergraph substitution rule:

\begin{equation}
\left\lbrace \left\lbrace x, y \right\rbrace, \left\lbrace x, z \right\rbrace \right\rbrace \to \left\lbrace \left\lbrace x, z \right\rbrace, \left\lbrace x, w \right\rbrace, \left\lbrace y, w \right\rbrace \right\rbrace.
\end{equation}
In Figure \ref{fig:Figure3}, we show the corresponding proof graph for this path, with pointed light green boxes representing axioms, dark orange triangles representing critical pair lemmas, light orange circles representing substitution lemmas, dark green diamonds representing hypotheses, solid lines representing substitutions, and dashed lines representing derived inference rules.

In this proof graph, we are effectively exploiting the causal structure ${\otimes_C}$ of ${\textbf{MuCauGraph}}$ in order to construct a refutation-complete proof calculus for first-order diagrammatic logic (with equality) that extends the conventional (unfailing) Knuth-Bendix completion procedure\cite{Knuth1983}, building upon the methods developed by Bachmair and Ganzinger\cite{Bachmair1994}. Specifically, we order equational terms using a selection function $S$ based upon the number of outgoing causal edges in the corresponding multiway evolution path, introducing deductive inference rules of \textit{selective resolution}:

\begin{equation}
\infer{\Lambda \sigma \implies \Pi \sigma}{\Lambda \cup \left\lbrace u \approx v \right\rbrace \implies \Pi},
\end{equation}
\textit{selective superposition}:

\begin{equation}
\infer{\left\lbrace u \left[ t \right] \sigma \approx v \sigma \right\rbrace \cup \Gamma \sigma \cup \Lambda \sigma \implies \Delta \sigma \cup \Pi \sigma}{\Gamma \implies \Delta \cup \left\lbrace s \approx t \right\rbrace \qquad \left\lbrace u \left[ s^{\prime} \right] \approx v \right\rbrace \cup \Lambda \implies \Pi},
\end{equation}
and \textit{ordered resolution}:

\begin{equation}
\infer{\Gamma \sigma \cup \Lambda \sigma \implies \Delta \sigma \cup \Pi \sigma}{\Gamma \implies \Delta \cup \left\lbrace P \left( s_1, \dots, s_n \right) \approx tt \right\rbrace \qquad \left\lbrace P \left( t_1, \dots, t_n \right) \approx tt \right\rbrace \cup \Lambda \implies \Pi},
\end{equation}
where ${u \approx v}$ is any occurrence of an equation within the clause:

\begin{equation}
\left\lbrace u \approx v \right\rbrace \cup \Lambda \implies \Pi,
\end{equation}
that is maximal with respect to the (causal) selection function $S$. Thus, when selecting which lemmas to add to the rewriting system (either substitution lemmas arising from instances of resolution/factoring, or critical pair lemmas arising from instances of completion/superposition/paramodulation), rather than using the standard reduction ordering on terms as in traditional unfailing completion approaches, we opt instead to select those lemmas which maximize the number of outgoing causal edges in the associated path of the multiway evolution causal graph, since intuitively this constitutes a reasonable heuristic for the lemmas which will exhibit the maximum effect on the shortening of subsequent proofs. The complete proof calculus for this reasoning system is presented in \cite{Gorard2021a}.

Although the example proof presented above considers the case of \textit{closed hypergraphs}, in which there are no ``dangling'' hyperedges, it is worth noting that both open and closed hypergraphs are ultimately special cases of the full typed hypergraph formalism presented within the preceding section. Specifically, for every hypergraph $G$, there exists a distinct hypergraph ${TG}$ (the \textit{type hypergraph}), with a total hypergraph morphism ${type_{G}}$ (the \textit{typing morphism}):

\begin{equation}
type_{G} : G \to TG.
\end{equation}
The following type graph, henceforth denoted ${2_{\mathcal{G}}}$, can be used to distinguish between ``true'' vertices and ``dummy'' vertices (which lie on the open ends of ``dangling'' hyperedges):

\begin{equation}
\begin{tikzcd}
V \arrow[r, bend right] & \epsilon \arrow[l, bend right] \arrow[loop right]
\end{tikzcd}.
\end{equation}
The category of open hypergraph string diagrams ${\textbf{OHGraph}}$ is therefore given by a subcategory of the slice category ${\left( \textbf{HGraph} \downarrow 2_{\mathcal{G}} \right)}$, where ${\textbf{HGraph}}$ designates the category of closed hypergraph string diagrams\cite{Dixon2013}.

\section{An Application to Quantum Circuit Simplification in the ZX-Calculus}
\label{sec:Section3}

Since open graph string diagrams, such as ZX-diagrams, constitute an important special case of the general (open/closed) hypergraph string diagrams discussed above, we proceed to illustrate how this formalism may be specialized to the case of the ZX-calculus. We choose to represent the diagrammatic rewriting rules of the ZX-calculus, of which many (such as the Z- and X-spider fusion rules) are in actual fact infinite rule schemas - with one rule representing each possible choice of spider arity - as rules in a second-order diagrammatic logic. From here, a tedious but straightforward construction due to Kerber\cite{Kerber1991a} allows us to define a morphism ${\Theta}$ from higher-order diagrammatic logics ${\mathcal{L}^{n}}$ to the first-order multisorted logic (with equality) ${\mathcal{L}_{sort}^{1}}$:

\begin{equation}
\Theta : \mathcal{L}^n \left( S \right) \to \mathcal{L}_{sort}^{1} \left( \Theta \left( S \right) \right),
\end{equation}
with $S$ and ${\Theta \left( S \right)}$ denoting the signatures of the logics ${\mathcal{L}^{n}}$ and ${\mathcal{L}_{sort}^{1}}$, respectively, such that sets of formulas ${\mathcal{F}}$ in the higher-order logic ${\mathcal{L}^{n} \left( S \right)}$ are always mapped to sets of formulas in the first-order multisorted logic ${\mathcal{L}_{sort}^{1} \left( \Theta \left( S \right) \right)}$:

\begin{equation}
\Theta : \mathcal{F} \left( \mathcal{L}^{n} \left( S \right) \right) \to \mathcal{F} \left( \mathcal{L}_{sort}^{1}  \left( \Theta \left( S \right) \right) \right).
\end{equation}
The complete translation between logics is presented in \cite{Gorard2021a}.

\begin{figure}[ht]
\centering
\includegraphics[align=c, width=0.295\textwidth]{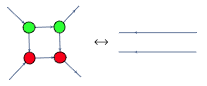}\hspace{0.1\textwidth}
\includegraphics[align=c, width=0.595\textwidth]{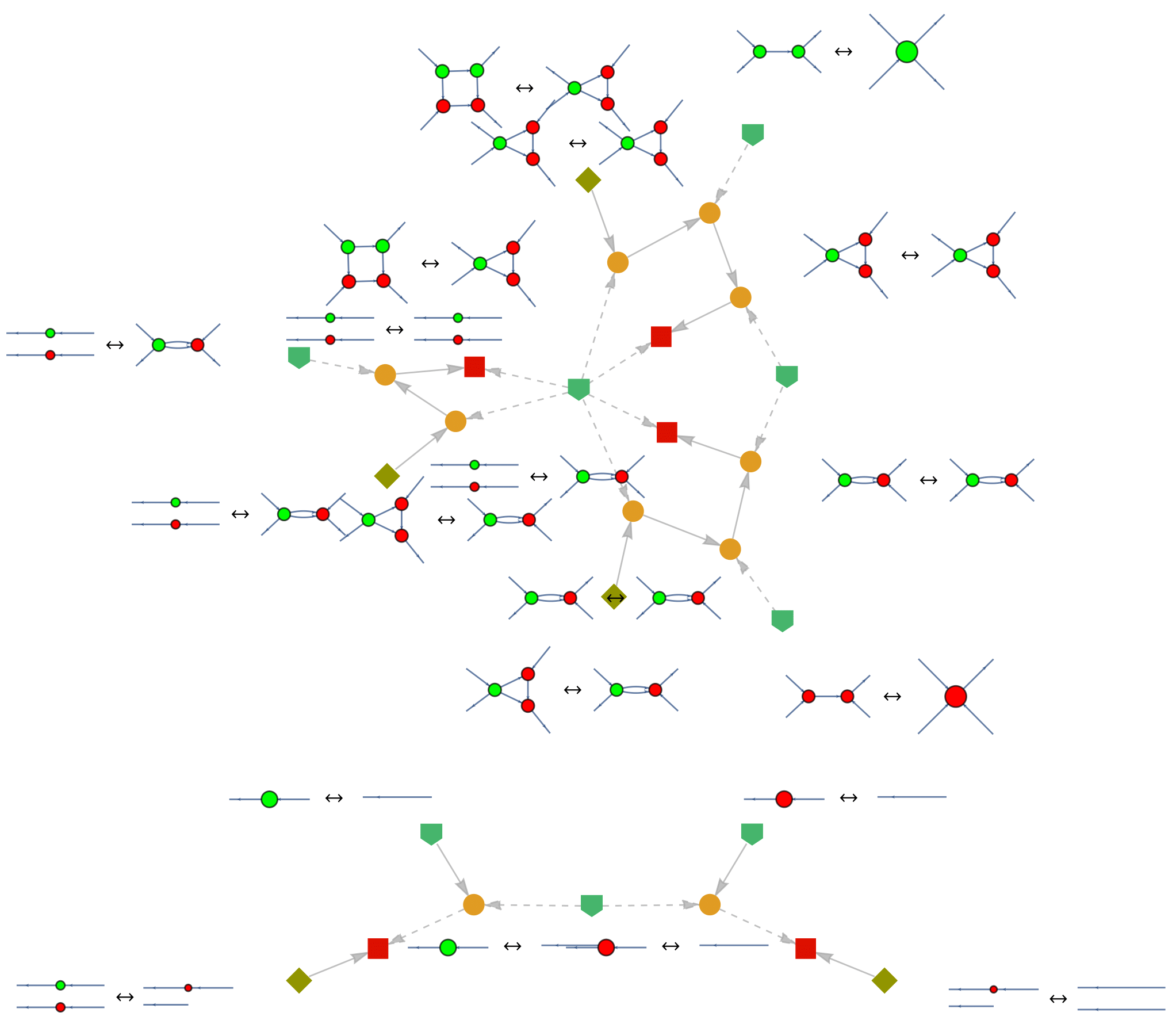}
\caption{On the left, the statement of unitarity of the CNOT gate, represented as a diagrammatic equality theorem in the ZX-calculus. On the right, the corresponding proof graph for this theorem.}
\label{fig:Figure4}
\end{figure}

We now proceed to demonstrate how this automatic diagrammatic reasoning system can be applied to the rather practical problem of the optimization of quantum circuits, by constructing an automated proof of unitarity of the CNOT gate ${\wedge X}$ in the ZX-calculus\cite{Coecke2009a}. Each CNOT gate ${\wedge X}$ is a simple two-spider ZX-diagram (consisting of a single Z-spider and a single X-spider), so the composition of two CNOT gates ${\wedge X \circ \wedge X}$ forms a four-spider diagram. The statement that the CNOT gate is unitary can therefore be represented as a diagrammatic equality between this four-spider diagram and a pair of identity morphisms ${1_{Q}^{\otimes 2}}$ (represented as a pair of parallel wires); this theorem, together with its associated proof graph, is shown in Figure \ref{fig:Figure4}.

\begin{figure}[ht]
\centering
\includegraphics[align=c, width=0.295\textwidth]{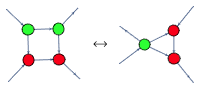}\hspace{0.1\textwidth}
\includegraphics[align=c, width=0.495\textwidth]{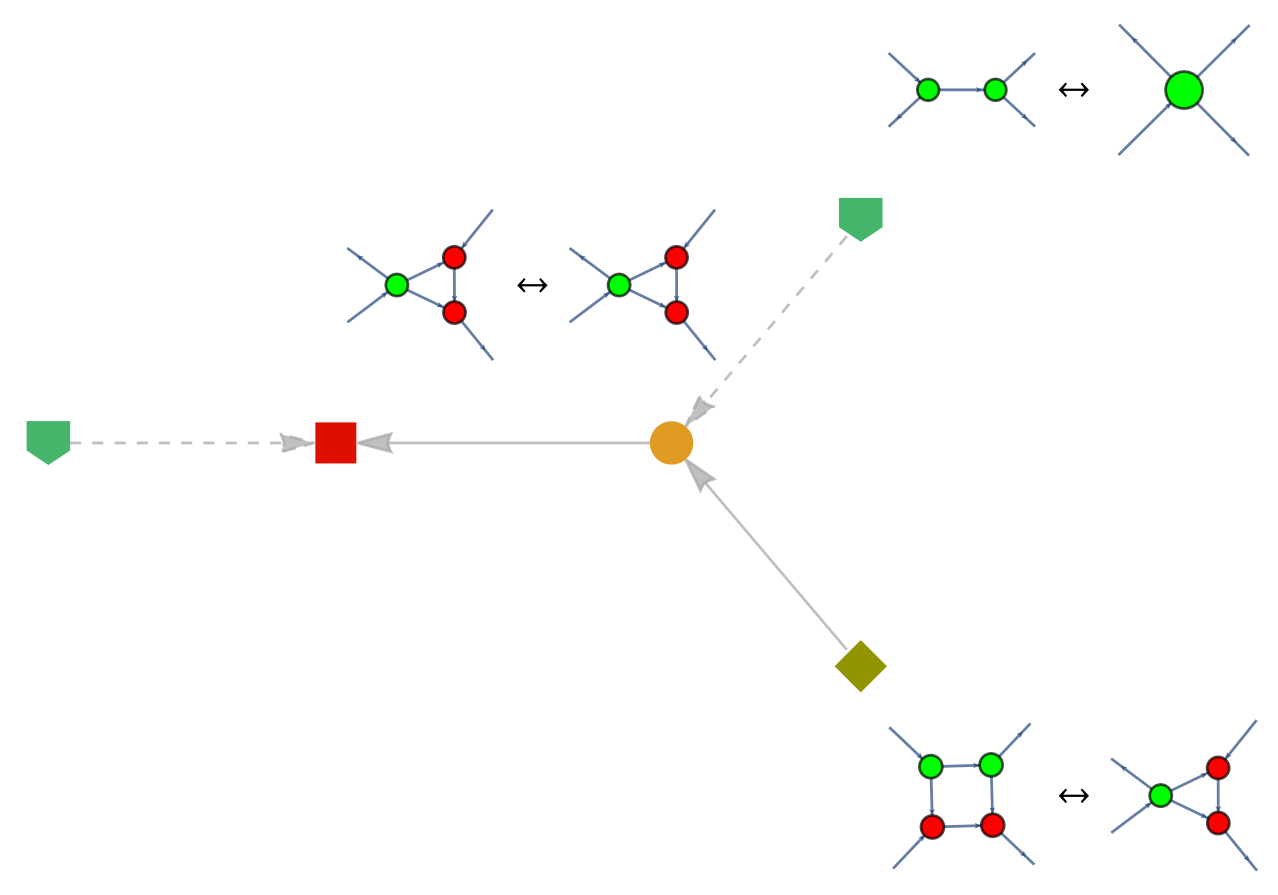}
\caption{On the left, the diagrammatic equality obtained by application of the Z-spider fusion rule. On the right, the corresponding proof graph for this deduction.}
\label{fig:Figure5}
\end{figure}

\begin{figure}[ht]
\centering
\includegraphics[align=c, width=0.295\textwidth]{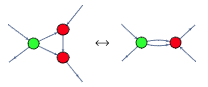}\hspace{0.1\textwidth}
\includegraphics[align=c, width=0.495\textwidth]{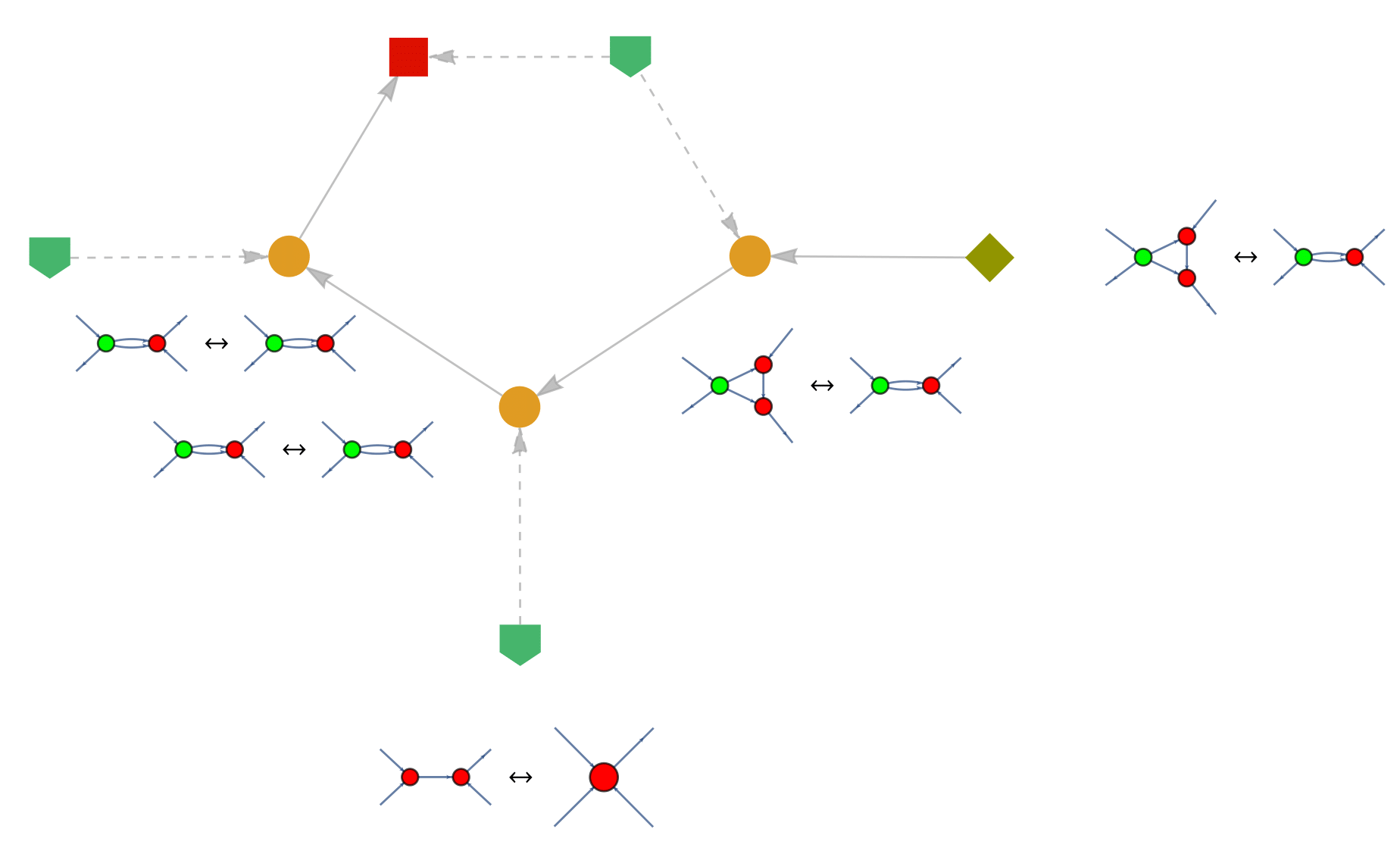}
\caption{On the left, the diagrammatic equality obtained by application of the X-spider fusion rule. On the right, the corresponding proof graph for this deduction.}
\label{fig:Figure6}
\end{figure}

However, we can see in detail how this proof is constructed by first applying the Z-spider fusion rule (S1) to the initial diagram in order to fuse the two phaseless Z-spiders from the two initial ${\wedge X}$ gates together, yielding the lemma shown in Figure \ref{fig:Figure5}, and likewise applying the X-spider fusion rule (S1) in order to fuse the two phaseless X-spiders from the two initial ${\wedge X}$ gates in much the same way, yielding the lemma shown in Figure \ref{fig:Figure6}.

\begin{figure}[ht]
\centering
\includegraphics[align=c, width=0.295\textwidth]{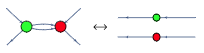}\hspace{0.1\textwidth}
\includegraphics[align=c, width=0.495\textwidth]{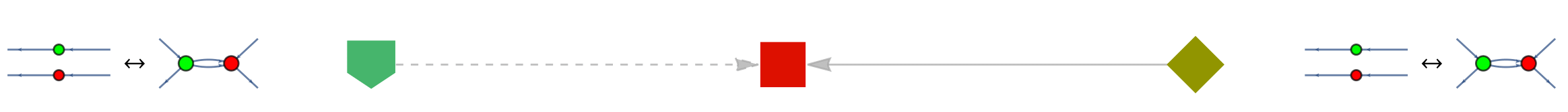}
\caption{On the left, the diagrammatic equality obtained by application of the Hopf law. On the right, the corresponding proof graph for this deduction.}
\label{fig:Figure7}
\end{figure}

Now, we must make use of the derived inference rule ${B^{\prime}}$ (otherwise known as the \textit{Hopf law}), which is deduced by a combination of the B1 (copy), B2 (bialgebra simplification), D2 (diamond) and S (fusion/identity) rules. When combined with the bialgebra simplification rule B2, the \textit{Hopf law} corresponds to the statement that interactions between spiders with differing colors in general yield \textit{scaled bialgebras} (i.e. structures that differ from bialgebras only by the presence of a normalizing factor). Thus, from here, we proceed to apply the Hopf law ${B^{\prime}}$ in order to ``untangle'' the Z- and X-spiders to yield a pair of parallel wires, each with a single Z/X-spider on it, as shown in Figure \ref{fig:Figure7}.

\begin{figure}[ht]
\centering
\includegraphics[align=c, width=0.295\textwidth]{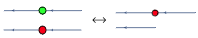}\hspace{0.1\textwidth}
\includegraphics[align=c, width=0.495\textwidth]{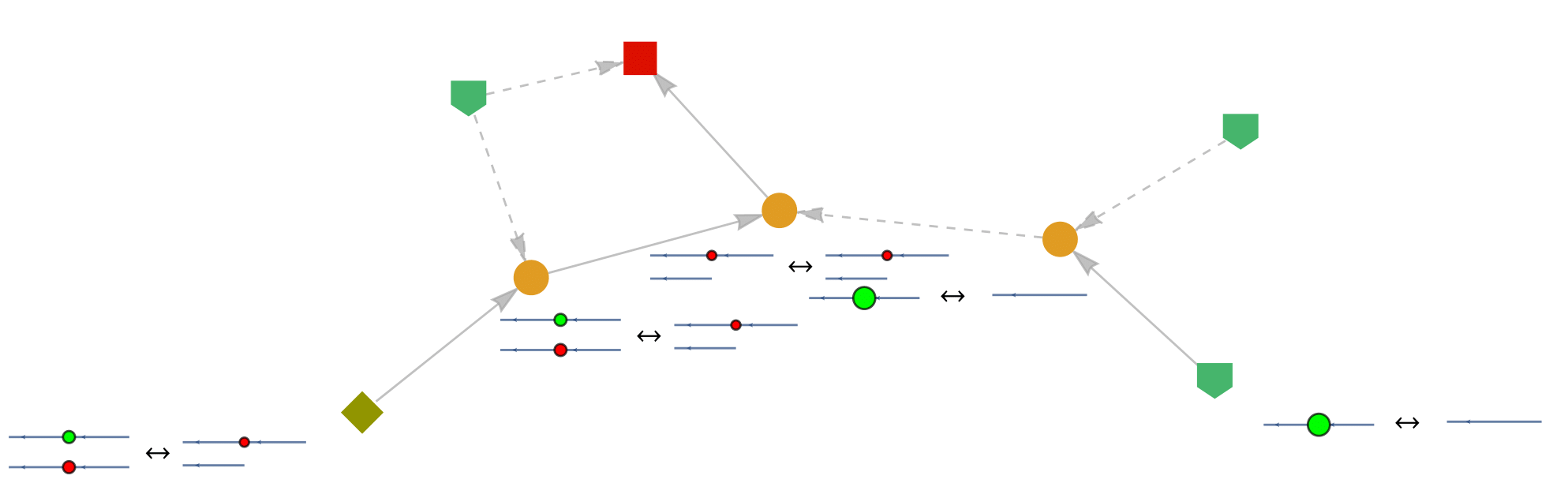}
\caption{On the left, the diagrammatic equality obtained by application of the Z-spider identity rule. On the right, the corresponding proof graph for this deduction.}
\label{fig:Figure8}
\end{figure}

\begin{figure}[ht]
\centering
\includegraphics[align=c, width=0.295\textwidth]{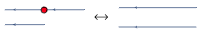}\hspace{0.1\textwidth}
\includegraphics[align=c, width=0.495\textwidth]{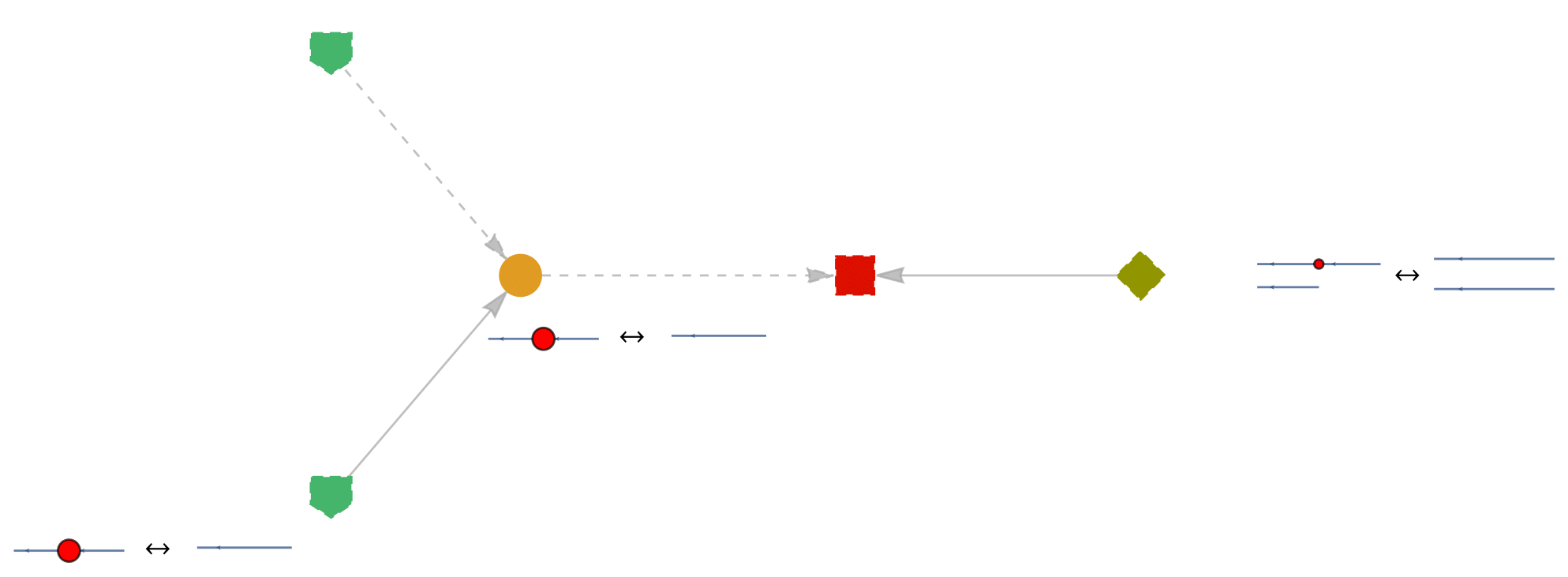}
\caption{On the left, the diagrammatic equality obtained by application of the X-spider identity rule. On the right, the corresponding proof graph for this deduction.}
\label{fig:Figure9}
\end{figure}

Finally, we apply the Z-spider identity rule (S2), followed by the X-spider identity rule (S2), in order to remove the remaining phaseless Z- and X-spiders from the diagram entirely, replacing them with single (parallel) wires, as required. These final two lemmas are shown in Figures \ref{fig:Figure8} and \ref{fig:Figure9}, respectively.

A complete performance analysis of this automated rewriting framework for the case of simplification of randomly-generated Clifford circuits to pseudo normal form and reduction of T-gates in randomly-generated non-Clifford circuits, both with and without causal structure optimization, and comparing across both proof complexity and time complexity metrics, is presented in \cite{Gorard2021a}. The results are found to be consistent with a roughly quadratic speedup across all performance metrics when causal structure optimization is applied.

\clearpage

\section*{Acknowledgments}

The authors would like to thank Nicolas Behr, Bob Coecke, Ross Duncan, Hatem Elshatlawy, Fabrizio Genovese, Reiko Heckel, Aleks Kissinger, David I. Spivak and Stephen Wolfram for useful conversations and/or suggestions. The authors would also like to thank the Foundations, Structures and Quantum Group at the University of Oxford, Cambridge Quantum Computing, the GReTA seminar series at the University of Paris, and the Topos Institute, for graciously inviting us to present various intermediate stages of this research and receive useful feedback.

{\footnotesize
\bibliographystyle{eptcs}
\bibliography{ACTBibliographyNew.bib}}

\begin{thebibliography}{10}
\providecommand{\bibitemdeclare}[2]{}
\providecommand{\surnamestart}{}
\providecommand{\surnameend}{}
\providecommand{\urlprefix}{Available at }
\providecommand{\url}[1]{\texttt{#1}}
\providecommand{\href}[2]{\texttt{#2}}
\providecommand{\urlalt}[2]{\href{#1}{#2}}
\providecommand{\doi}[1]{doi:\urlalt{http://dx.doi.org/#1}{#1}}
\providecommand{\bibinfo}[2]{#2}

\bibitemdeclare{inproceedings}{Abramsky2004}
\bibitem{Abramsky2004}
\bibinfo{author}{Samson \surnamestart Abramsky\surnameend} \&
  \bibinfo{author}{Bob \surnamestart Coecke\surnameend} (\bibinfo{year}{2004}):
  \emph{\bibinfo{title}{A Categorical Semantics of Quantum Protocols}}.
\newblock In: {\sl \bibinfo{booktitle}{19th {IEEE} Symposium on Logic in
  Computer Science {(LICS} 2004), 14-17 July 2004, Turku, Finland,
  Proceedings}}, \bibinfo{publisher}{{IEEE} Computer Society}, pp.
  \bibinfo{pages}{415--425}, \doi{10.1109/LICS.2004.1319636}.

\bibitemdeclare{incollection}{Abramsky2009}
\bibitem{Abramsky2009}
\bibinfo{author}{Samson \surnamestart {Abramsky}\surnameend} \&
  \bibinfo{author}{Bob \surnamestart {Coecke}\surnameend}
  (\bibinfo{year}{2009}): \emph{\bibinfo{title}{Categorical quantum
  mechanics}}.
\newblock In: {\sl \bibinfo{booktitle}{{Handbook of quantum logic and quantum
  structures. Quantum logic. With a foreword by Anatolij Dvure\v{c}enskij}}},
  \bibinfo{publisher}{Amsterdam: Elsevier/North-Holland}, pp.
  \bibinfo{pages}{261--323}, \doi{10.1016/b978-0-444-52869-8.50010-4}.

\bibitemdeclare{article}{Bachmair1994}
\bibitem{Bachmair1994}
\bibinfo{author}{Leo \surnamestart {Bachmair}\surnameend} \&
  \bibinfo{author}{Harald \surnamestart {Ganzinger}\surnameend}
  (\bibinfo{year}{1994}): \emph{\bibinfo{title}{Rewrite-based equational
  theorem proving with selection and simplification}}.
\newblock {\sl \bibinfo{journal}{Journal of Logic and Computation}}
  \bibinfo{volume}{4}(\bibinfo{number}{3}), pp. \bibinfo{pages}{217--247},
  \doi{10.1093/logcom/4.3.217}.

\bibitemdeclare{article}{Baez2014}
\bibitem{Baez2014}
\bibinfo{author}{John~C. \surnamestart Baez\surnameend} \&
  \bibinfo{author}{Jason \surnamestart Erbele\surnameend}
  (\bibinfo{year}{2014}): \emph{\bibinfo{title}{Categories in Control}}.
\newblock {\sl \bibinfo{journal}{Theory and Applications of Categories, Vol. 30
  No. 24 (2015), 836-881}}.

\bibitemdeclare{article}{Baez2018}
\bibitem{Baez2018}
\bibinfo{author}{John~C. \surnamestart {Baez}\surnameend} \&
  \bibinfo{author}{Brendan \surnamestart {Fong}\surnameend}
  (\bibinfo{year}{2018}): \emph{\bibinfo{title}{A compositional framework for
  passive linear networks}}.
\newblock {\sl \bibinfo{journal}{Theory and Applications of Categories}}
  \bibinfo{volume}{33}, pp. \bibinfo{pages}{1158--1222}.

\bibitemdeclare{article}{Baez2017}
\bibitem{Baez2017}
\bibinfo{author}{John~C. \surnamestart {Baez}\surnameend} \&
  \bibinfo{author}{Blake~S. \surnamestart {Pollard}\surnameend}
  (\bibinfo{year}{2017}): \emph{\bibinfo{title}{A compositional framework for
  reaction networks}}.
\newblock {\sl \bibinfo{journal}{Reviews in Mathematical Physics}}
  \bibinfo{volume}{29}(\bibinfo{number}{9}), p.~\bibinfo{pages}{41},
  \doi{10.1142/s0129055x17500283}.
\newblock \bibinfo{note}{Id/No 1750028}.

\bibitemdeclare{incollection}{Benabou1967}
\bibitem{Benabou1967}
\bibinfo{author}{Jean \surnamestart {B\'enabou}\surnameend}
  (\bibinfo{year}{1967}): \emph{\bibinfo{title}{Introduction to bicategories}}.
\newblock In: {\sl \bibinfo{booktitle}{{Reports of the Midwest Category
  Seminar}}}, pp. \bibinfo{pages}{1--77}, \doi{10.1007/bfb0074299}.

\bibitemdeclare{article}{Bolt2017}
\bibitem{Bolt2017}
\bibinfo{author}{Joe \surnamestart Bolt\surnameend}, \bibinfo{author}{Bob
  \surnamestart Coecke\surnameend}, \bibinfo{author}{Fabrizio \surnamestart
  Genovese\surnameend}, \bibinfo{author}{Martha \surnamestart
  Lewis\surnameend}, \bibinfo{author}{Dan \surnamestart Marsden\surnameend} \&
  \bibinfo{author}{Robin \surnamestart Piedeleu\surnameend}
  (\bibinfo{year}{2017}): \emph{\bibinfo{title}{Interacting Conceptual Spaces
  {I} : Grammatical Composition of Concepts}}.
\newblock {\sl \bibinfo{journal}{CoRR}} \bibinfo{volume}{abs/1703.08314},
  \doi{10.1007/978-3-030-12800-5\_9}.
\newblock \urlprefix\url{http://arxiv.org/abs/1703.08314}.

\bibitemdeclare{inproceedings}{Bonchi2014}
\bibitem{Bonchi2014}
\bibinfo{author}{Filippo \surnamestart Bonchi\surnameend},
  \bibinfo{author}{Pawel \surnamestart Sobocinski\surnameend} \&
  \bibinfo{author}{Fabio \surnamestart Zanasi\surnameend}
  (\bibinfo{year}{2014}): \emph{\bibinfo{title}{A Categorical Semantics of
  Signal Flow Graphs}}.
\newblock In \bibinfo{editor}{Paolo \surnamestart Baldan\surnameend} \&
  \bibinfo{editor}{Daniele \surnamestart Gorla\surnameend}, editors: {\sl
  \bibinfo{booktitle}{{CONCUR} 2014 - Concurrency Theory - 25th International
  Conference, {CONCUR} 2014, Rome, Italy, September 2-5, 2014. Proceedings}},
  {\sl \bibinfo{series}{Lecture Notes in Computer Science}}
  \bibinfo{volume}{8704}, \bibinfo{publisher}{Springer}, pp.
  \bibinfo{pages}{435--450}, \doi{10.1007/978-3-662-44584-6\_30}.
\newblock \urlprefix\url{https://doi.org/10.1007/978-3-662-44584-6_30}.

\bibitemdeclare{inproceedings}{Coecke2008}
\bibitem{Coecke2008}
\bibinfo{author}{Bob \surnamestart Coecke\surnameend} \& \bibinfo{author}{Ross
  \surnamestart Duncan\surnameend} (\bibinfo{year}{2008}):
  \emph{\bibinfo{title}{Interacting Quantum Observables}}.
\newblock In \bibinfo{editor}{Luca \surnamestart Aceto\surnameend},
  \bibinfo{editor}{Ivan \surnamestart Damg{\aa}rd\surnameend},
  \bibinfo{editor}{Leslie~Ann \surnamestart Goldberg\surnameend},
  \bibinfo{editor}{Magn{\'{u}}s~M. \surnamestart Halld{\'{o}}rsson\surnameend},
  \bibinfo{editor}{Anna \surnamestart Ing{\'{o}}lfsd{\'{o}}ttir\surnameend} \&
  \bibinfo{editor}{Igor \surnamestart Walukiewicz\surnameend}, editors: {\sl
  \bibinfo{booktitle}{Automata, Languages and Programming, 35th International
  Colloquium, {ICALP} 2008, Reykjavik, Iceland, July 7-11, 2008, Proceedings,
  Part {II} - Track {B:} Logic, Semantics, and Theory of Programming {\&} Track
  {C:} Security and Cryptography Foundations}}, {\sl \bibinfo{series}{Lecture
  Notes in Computer Science}} \bibinfo{volume}{5126},
  \bibinfo{publisher}{Springer}, pp. \bibinfo{pages}{298--310},
  \doi{10.1007/978-3-540-70583-3\_25}.
\newblock \urlprefix\url{https://doi.org/10.1007/978-3-540-70583-3_25}.

\bibitemdeclare{article}{Coecke2009a}
\bibitem{Coecke2009a}
\bibinfo{author}{Bob \surnamestart Coecke\surnameend} \& \bibinfo{author}{Ross
  \surnamestart Duncan\surnameend} (\bibinfo{year}{2009}):
  \emph{\bibinfo{title}{Interacting Quantum Observables: Categorical Algebra
  and Diagrammatics}}.
\newblock {\sl \bibinfo{journal}{New J. Phys. 13 (2011) 043016}},
  \doi{10.1088/1367-2630/13/4/043016}.

\bibitemdeclare{article}{Coecke2013}
\bibitem{Coecke2013}
\bibinfo{author}{Bob \surnamestart Coecke\surnameend} \&
  \bibinfo{author}{Raymond \surnamestart Lal\surnameend}
  (\bibinfo{year}{2013}): \emph{\bibinfo{title}{Causal categories:
  relativistically interacting processes}}.
\newblock {\sl \bibinfo{journal}{Found. Phys.}} \bibinfo{volume}{43}, pp.
  \bibinfo{pages}{458--501}, \doi{10.1007/s10701-012-9646-8}.

\bibitemdeclare{article}{Coecke2010a}
\bibitem{Coecke2010a}
\bibinfo{author}{Bob \surnamestart Coecke\surnameend},
  \bibinfo{author}{Mehrnoosh \surnamestart Sadrzadeh\surnameend} \&
  \bibinfo{author}{Stephen \surnamestart Clark\surnameend}
  (\bibinfo{year}{2010}): \emph{\bibinfo{title}{Mathematical Foundations for a
  Compositional Distributional Model of Meaning}}.
\newblock {\sl \bibinfo{journal}{Lambek Festschirft, special issue of
  Linguistic Analysis, 2010.}}

\bibitemdeclare{article}{Dixon2013}
\bibitem{Dixon2013}
\bibinfo{author}{Lucas \surnamestart {Dixon}\surnameend} \&
  \bibinfo{author}{Aleks \surnamestart {Kissinger}\surnameend}
  (\bibinfo{year}{2013}): \emph{\bibinfo{title}{Open-graphs and monoidal
  theories}}.
\newblock {\sl \bibinfo{journal}{MSCS. Mathematical Structures in Computer
  Science}} \bibinfo{volume}{23}(\bibinfo{number}{2}), pp.
  \bibinfo{pages}{308--359}, \doi{10.1017/s0960129512000138}.

\bibitemdeclare{book}{Ehrig2006}
\bibitem{Ehrig2006}
\bibinfo{author}{Hartmut \surnamestart Ehrig\surnameend},
  \bibinfo{author}{Karsten \surnamestart Ehrig\surnameend},
  \bibinfo{author}{Ulrike \surnamestart Prange\surnameend} \&
  \bibinfo{author}{Gabriele \surnamestart Taentzer\surnameend}
  (\bibinfo{year}{2006}): \emph{\bibinfo{title}{Fundamentals of Algebraic Graph
  Transformation}}.
\newblock \bibinfo{series}{Monographs in Theoretical Computer Science. An
  {EATCS} Series}, \bibinfo{publisher}{Springer}, \doi{10.1007/3-540-31188-2}.

\bibitemdeclare{inproceedings}{Ehrig1973}
\bibitem{Ehrig1973}
\bibinfo{author}{Hartmut \surnamestart Ehrig\surnameend},
  \bibinfo{author}{Michael \surnamestart Pfender\surnameend} \&
  \bibinfo{author}{Hans~J{\"{u}}rgen \surnamestart Schneider\surnameend}
  (\bibinfo{year}{1973}): \emph{\bibinfo{title}{Graph-Grammars: An Algebraic
  Approach}}.
\newblock In: {\sl \bibinfo{booktitle}{14th Annual Symposium on Switching and
  Automata Theory, Iowa City, Iowa, USA, October 15-17, 1973}},
  \bibinfo{publisher}{{IEEE} Computer Society}, pp. \bibinfo{pages}{167--180},
  \doi{10.1109/SWAT.1973.11}.

\bibitemdeclare{article}{Fong2015}
\bibitem{Fong2015}
\bibinfo{author}{Brendan \surnamestart {Fong}\surnameend}
  (\bibinfo{year}{2015}): \emph{\bibinfo{title}{Decorated cospans}}.
\newblock {\sl \bibinfo{journal}{Theory and Applications of Categories}}
  \bibinfo{volume}{30}, pp. \bibinfo{pages}{1096--1120}.

\bibitemdeclare{phdthesis}{Fong2016}
\bibitem{Fong2016}
\bibinfo{author}{Brendan \surnamestart Fong\surnameend} (\bibinfo{year}{2016}):
  \emph{\bibinfo{title}{The algebra of open and interconnected systems}}.
\newblock Ph.D. thesis, \bibinfo{school}{University of Oxford, {UK}}.
\newblock
  \urlprefix\url{http://ethos.bl.uk/OrderDetails.do?uin=uk.bl.ethos.730061}.

\bibitemdeclare{article}{Gorard2016}
\bibitem{Gorard2016}
\bibinfo{author}{Jonathan \surnamestart Gorard\surnameend}
  (\bibinfo{year}{2016}): \emph{\bibinfo{title}{Uniqueness Trees: {A} Possible
  Polynomial Approach to the Graph Isomorphism Problem}}.
\newblock {\sl \bibinfo{journal}{CoRR}} \bibinfo{volume}{abs/1606.06399}.
\newblock \urlprefix\url{http://arxiv.org/abs/1606.06399}.

\bibitemdeclare{article}{Gorard2020c}
\bibitem{Gorard2020c}
\bibinfo{author}{Jonathan \surnamestart Gorard\surnameend}
  (\bibinfo{year}{2020}): \emph{\bibinfo{title}{Algorithmic Causal Sets and the
  Wolfram Model}}.
\newblock {\sl \bibinfo{journal}{CoRR}} \bibinfo{volume}{abs/2011.12174}.
\newblock \urlprefix\url{https://arxiv.org/abs/2011.12174}.

\bibitemdeclare{article}{Gorard2020}
\bibitem{Gorard2020}
\bibinfo{author}{Jonathan \surnamestart Gorard\surnameend}
  (\bibinfo{year}{2020}): \emph{\bibinfo{title}{Some Quantum Mechanical
  Properties of the Wolfram Model}}.
\newblock {\sl \bibinfo{journal}{Complex Syst.}}
  \bibinfo{volume}{29}(\bibinfo{number}{2}),
  \doi{10.25088/complexsystems.29.2.537}.
\newblock
  \urlprefix\url{https://www.complex-systems.com/abstracts/v29_i02_a02/}.

\bibitemdeclare{article}{Gorard2020d}
\bibitem{Gorard2020d}
\bibinfo{author}{Jonathan \surnamestart Gorard\surnameend}
  (\bibinfo{year}{2020}): \emph{\bibinfo{title}{Some Relativistic and
  Gravitational Properties of the Wolfram Model}}.
\newblock {\sl \bibinfo{journal}{Complex Syst.}}
  \bibinfo{volume}{29}(\bibinfo{number}{2}),
  \doi{10.25088/complexsystems.29.2.599}.
\newblock
  \urlprefix\url{https://www.complex-systems.com/abstracts/v29_i02_a03/}.

\bibitemdeclare{article}{Gorard2021b}
\bibitem{Gorard2021b}
\bibinfo{author}{Jonathan \surnamestart Gorard\surnameend}
  (\bibinfo{year}{2021}): \emph{\bibinfo{title}{Hypergraph Discretization of
  the Cauchy Problem in General Relativity via Wolfram Model Evolution}}.
\newblock \urlprefix\url{http://arxiv.org/pdf/2102.09363v1}.

\bibitemdeclare{article}{Gorard2020b}
\bibitem{Gorard2020b}
\bibinfo{author}{Jonathan \surnamestart Gorard\surnameend},
  \bibinfo{author}{Manojna \surnamestart Namuduri\surnameend} \&
  \bibinfo{author}{Xerxes~D. \surnamestart Arsiwalla\surnameend}
  (\bibinfo{year}{2020}): \emph{\bibinfo{title}{ZX-Calculus and Extended
  Hypergraph Rewriting Systems {I:} {A} Multiway Approach to Categorical
  Quantum Information Theory}}.
\newblock {\sl \bibinfo{journal}{CoRR}} \bibinfo{volume}{abs/2010.02752}.
\newblock \urlprefix\url{https://arxiv.org/abs/2010.02752}.

\bibitemdeclare{article}{Gorard2021a}
\bibitem{Gorard2021a}
\bibinfo{author}{Jonathan \surnamestart Gorard\surnameend},
  \bibinfo{author}{Manojna \surnamestart Namuduri\surnameend} \&
  \bibinfo{author}{Xerxes~D. \surnamestart Arsiwalla\surnameend}
  (\bibinfo{year}{2021}): \emph{\bibinfo{title}{ZX-Calculus and Extended
  Wolfram Model Systems {II:} Fast Diagrammatic Reasoning with an Application
  to Quantum Circuit Simplification}}.
\newblock {\sl \bibinfo{journal}{CoRR}} \bibinfo{volume}{abs/2103.15820}.
\newblock \urlprefix\url{https://arxiv.org/abs/2103.15820}.

\bibitemdeclare{article}{Joyal1991}
\bibitem{Joyal1991}
\bibinfo{author}{Andr\'e \surnamestart {Joyal}\surnameend} \&
  \bibinfo{author}{Ross \surnamestart {Street}\surnameend}
  (\bibinfo{year}{1991}): \emph{\bibinfo{title}{The geometry of tensor
  calculus. I}}.
\newblock {\sl \bibinfo{journal}{Advances in Mathematics}}
  \bibinfo{volume}{88}(\bibinfo{number}{1}), pp. \bibinfo{pages}{55--112},
  \doi{10.1016/0001-8708(91)90003-p}.

\bibitemdeclare{inproceedings}{Kerber1991a}
\bibitem{Kerber1991a}
\bibinfo{author}{Manfred \surnamestart Kerber\surnameend}
  (\bibinfo{year}{1991}): \emph{\bibinfo{title}{How to Prove Higher Order
  Theorems in First Order Logic}}.
\newblock In \bibinfo{editor}{John \surnamestart Mylopoulos\surnameend} \&
  \bibinfo{editor}{Raymond \surnamestart Reiter\surnameend}, editors: {\sl
  \bibinfo{booktitle}{Proceedings of the 12th International Joint Conference on
  Artificial Intelligence. Sydney, Australia, August 24-30, 1991}},
  \bibinfo{publisher}{Morgan Kaufmann}, pp. \bibinfo{pages}{137--142}.
\newblock \urlprefix\url{http://ijcai.org/Proceedings/91-1/Papers/023.pdf}.

\bibitemdeclare{phdthesis}{Kissinger2011}
\bibitem{Kissinger2011}
\bibinfo{author}{Aleks \surnamestart Kissinger\surnameend}
  (\bibinfo{year}{2011}): \emph{\bibinfo{title}{Pictures of processes :
  automated graph rewriting for monoidal categories and applications to quantum
  computing}}.
\newblock Ph.D. thesis, \bibinfo{school}{University of Oxford, {UK}}.
\newblock
  \urlprefix\url{http://ora.ox.ac.uk/objects/uuid:61fb3161-a353-48fc-8da2-6ce220cce6a2}.

\bibitemdeclare{article}{Kissinger2014}
\bibitem{Kissinger2014}
\bibinfo{author}{Aleks \surnamestart Kissinger\surnameend}
  (\bibinfo{year}{2014}): \emph{\bibinfo{title}{Finite matrices are complete
  for (dagger-)hypergraph categories}}.

\bibitemdeclare{article}{Kissinger2019}
\bibitem{Kissinger2019}
\bibinfo{author}{Aleks \surnamestart Kissinger\surnameend} \&
  \bibinfo{author}{John \surnamestart van~de Wetering\surnameend}
  (\bibinfo{year}{2019}): \emph{\bibinfo{title}{PyZX: Large Scale Automated
  Diagrammatic Reasoning}}.
\newblock {\sl \bibinfo{journal}{EPTCS 318, 2020, pp. 229-241}},
  \doi{10.4204/EPTCS.318.14}.

\bibitemdeclare{inproceedings}{Kissinger2015}
\bibitem{Kissinger2015}
\bibinfo{author}{Aleks \surnamestart Kissinger\surnameend} \&
  \bibinfo{author}{Vladimir \surnamestart Zamdzhiev\surnameend}
  (\bibinfo{year}{2015}): \emph{\bibinfo{title}{Quantomatic: {A} Proof
  Assistant for Diagrammatic Reasoning}}.
\newblock In \bibinfo{editor}{Amy~P. \surnamestart Felty\surnameend} \&
  \bibinfo{editor}{Aart \surnamestart Middeldorp\surnameend}, editors: {\sl
  \bibinfo{booktitle}{Automated Deduction - {CADE-25} - 25th International
  Conference on Automated Deduction, Berlin, Germany, August 1-7, 2015,
  Proceedings}}, {\sl \bibinfo{series}{Lecture Notes in Computer Science}}
  \bibinfo{volume}{9195}, \bibinfo{publisher}{Springer}, pp.
  \bibinfo{pages}{326--336}, \doi{10.1007/978-3-319-21401-6\_22}.
\newblock \urlprefix\url{https://doi.org/10.1007/978-3-319-21401-6_22}.

\bibitemdeclare{misc}{Knuth1983}
\bibitem{Knuth1983}
\bibinfo{author}{D.~E. \surnamestart Knuth\surnameend} \&
  \bibinfo{author}{P.~B. \surnamestart Bendix\surnameend}
  (\bibinfo{year}{1983}): \emph{\bibinfo{title}{Simple Word Problems in
  Universal Algebras}}, \doi{10.1007/978-3-642-81955-1\_23}.

\bibitemdeclare{inproceedings}{Lack2004a}
\bibitem{Lack2004a}
\bibinfo{author}{Stephen \surnamestart Lack\surnameend} \&
  \bibinfo{author}{Pawel \surnamestart Sobocinski\surnameend}
  (\bibinfo{year}{2004}): \emph{\bibinfo{title}{Adhesive Categories}}.
\newblock In \bibinfo{editor}{Igor \surnamestart Walukiewicz\surnameend},
  editor: {\sl \bibinfo{booktitle}{Foundations of Software Science and
  Computation Structures, 7th International Conference, {FOSSACS} 2004, Held as
  Part of the Joint European Conferences on Theory and Practice of Software,
  {ETAPS} 2004, Barcelona, Spain, March 29 - April 2, 2004, Proceedings}}, {\sl
  \bibinfo{series}{Lecture Notes in Computer Science}} \bibinfo{volume}{2987},
  \bibinfo{publisher}{Springer}, pp. \bibinfo{pages}{273--288},
  \doi{10.1007/978-3-540-24727-2\_20}.
\newblock \urlprefix\url{https://doi.org/10.1007/978-3-540-24727-2_20}.

\bibitemdeclare{book}{Wolfram2002a}
\bibitem{Wolfram2002a}
\bibinfo{author}{Stephen \surnamestart Wolfram\surnameend}
  (\bibinfo{year}{2002}): \emph{\bibinfo{title}{A new kind of science}}.
\newblock \bibinfo{publisher}{Wolfram Media}, \bibinfo{address}{Champaign,
  Ill.}

\bibitemdeclare{article}{Wolfram2020}
\bibitem{Wolfram2020}
\bibinfo{author}{Stephen \surnamestart Wolfram\surnameend}
  (\bibinfo{year}{2020}): \emph{\bibinfo{title}{A Class of Models with the
  Potential to Represent Fundamental Physics}}.
\newblock {\sl \bibinfo{journal}{Complex Syst.}} \bibinfo{volume}{29}, pp.
  \bibinfo{pages}{107--536}, \doi{10.25088/ComplexSystems.29.2.107}.

\end{thebibliography}

\end{document}